\documentclass[acmsmall]{acmart}
\usepackage{makecell}
\usepackage{enumitem}
\usepackage{graphicx}
\usepackage{mdframed}
\usepackage{amsmath}
\usepackage{tcolorbox}
\usepackage{multirow}
\usepackage[flushleft]{threeparttable}
\usepackage{tabularx,ragged2e}
\usepackage{subfig}
\usepackage{xcolor}
\usepackage{colortbl}
\usepackage{listings}
\newcolumntype{C}{>{\arraybackslash}X} 
\definecolor{blue}{HTML}{000000}
\definecolor{violet}{HTML}{000000}
\definecolor{teal}{HTML}{000000}
\definecolor{cyan}{HTML}{000000}
\definecolor{major-rev}{HTML}{1206CC}
\AtBeginDocument{%
  }

\setcopyright{acmlicensed}
\copyrightyear{2024}
\acmYear{2024}
\acmDOI{XXXXXXX.XXXXXXX}

\acmMonth{8}

\begin{document}

\title{Beyond ChatGPT: Enhancing Software Quality Assurance Tasks with Diverse LLMs and Validation Techniques}


\author{Ratnadira Widyasari}
\affiliation{%
  \institution{Singapore Management University}
  \city{Singapore}
  \country{Singapore}}
\email{ratnadiraw.2020@phdcs.smu.edu.sg}

\author{David Lo}
\affiliation{%
  \institution{Singapore Management University}
  \city{Singapore}
  \country{Singapore}}
\email{davidlo@smu.edu.sg}

\author{Lizi Liao}
\affiliation{%
  \institution{Singapore Management University}
  \city{Singapore}
  \country{Singapore}}
\email{lzliao@smu.edu.sg}


\begin{abstract}
With the advancement of Large Language Models (LLMs), their application in Software Quality Assurance (SQA) has increased. However, the current focus of these applications is predominantly on ChatGPT. There remains a gap in understanding the performance of various LLMs in this critical domain. 
This paper aims to address this gap by conducting a comprehensive investigation into the capabilities of several LLMs across two SQA tasks: fault localization and vulnerability detection.  We conducted comparative studies using GPT-3.5, GPT-4o, and four other prominent publicly available LLMs—LLaMA-3-70B, LLaMA-3-8B, Gemma-7B, and Mixtral-8x7B, to evaluate their effectiveness in these tasks.

Our findings reveal that several LLMs can outperform GPT-3.5 in both tasks. Additionally, even the lower-performing LLMs provided unique correct predictions, suggesting the potential of combining different LLMs' results to enhance overall performance. By implementing a voting mechanism to combine the LLMs' results, we achieved more than a 10\% improvement over the GPT-3.5 in both tasks. Furthermore, we introduced a cross-validation approach to refine the LLM answer by validating one LLM answer against another using a validation prompt. This approach led to performance improvements of 16\% in fault localization and 12\% in vulnerability detection compared to the GPT-3.5, with 4\% improvement compared to the best-performed LLMs. Our analysis also indicates that the inclusion of explanations in the LLMs' results affects the effectiveness of the cross-validation technique.
\end{abstract}

\ccsdesc[300]{Software and its engineering~Software maintenance tools}

\keywords{Software quality assurance, fault localization, vulnerability detection, LLM, GPT-4o, GPT-3.5, Gemma-7B, LLaMA-3, Mixtral}

\received{31 August 2024}

\maketitle

\section{Introduction}\label{sec:intro}
In recent years, Large Language Models (LLMs) have gained widespread popularity for their remarkable ability to generate natural and coherent text. Their application in the Software Quality Assurance (SQA) domain has yielded promising outcomes across various tasks, including automated program repair~\cite{xia2023automated, liu2023refining}, code review~\cite{rasheed2024aipowered}, vulnerability detection~\cite{zhou2024large,zhang2024prompt, fu2023chatgpt, cheshkov2023evaluation,tamberg2024harnessing}, and fault localization~\cite{wu2023large,kang2023preliminary,widyasari2024demystifying, jiang2024evaluating}. 

While these advancements are notable, most studies in this domain~\cite{zhou2024large,zhang2024prompt, fu2023chatgpt, cheshkov2023evaluation,wu2023large,kang2023preliminary,widyasari2024demystifying, jiang2024evaluating} have predominantly utilized GPT-3.5 from ChatGPT~\cite{OpenAI_Models} as the underlying LLM engine. There are, however, several other LLMs available aside from ChatGPT, such as Mixtral~\cite{Mistral_Mixtral}, Gemma~\cite{Google_Gemma}, which may offer unique advantages in SQA tasks. The focus of many previous studies on ChatGPT leaves an underexplored area regarding the potential advantages or limitations of employing a diverse range of LLMs in SQA tasks. 
The motivation behind this study lies in the hypothesis that not all LLMs are equally effective across different SQA tasks. For instance, while GPT-3.5 might excel in natural language generation, it may not necessarily be the best model for detecting subtle software vulnerabilities. 
Consider a scenario where an LLM is used exclusively for vulnerability detection, but it fails to identify the vulnerability due to inherent model biases. In contrast, a more diverse approach using multiple LLMs could mitigate these biases by leveraging the strengths of different models, leading to more accurate results.

To address this gap, we propose a comprehensive investigation that utilizes a diverse set of LLMs in SQA tasks, specifically in fault localization and vulnerability detection. Fault localization and vulnerability detection were chosen as the focus of this study due to their direct impact on software reliability and security~\cite{zou2019empirical}. Previous studies~\cite{wu2023large,kang2023preliminary, widyasari2024demystifying, zhou2024large,zhang2024prompt, fu2023chatgpt, cheshkov2023evaluation} that work on these two tasks have predominantly utilized ChatGPT as the LLM-based engine, which provides a substantial basis for comparison and improvement. Fault localization involves identifying the specific code location that causes software failure~\cite{pearson2016evaluating}, which is a crucial step in the debugging process.  
On the other hand, vulnerability detection is essential for identifying potential security weaknesses in the code~\cite{fu2023chatgpt}.

We conducted an analysis on various LLM results, including GPT-3.5~\cite{OpenAI_Models}, GPT-4o~\cite{OpenAI_Models},  LLaMA-3-70B~\cite{Meta_LLaMA3}, LLaMA-3-8B~\cite{Meta_LLaMA3}, Gemma-7B~\cite{Google_Gemma}, and Mixtral-8x7B~\cite{Mistral_Mixtral} (cf. Section~\ref{sec:llm_setting}). We used GPT-3.5 as a baseline, given its widespread use in previous studies, alongside state-of-the-art non-LLM-based techniques for both tasks. Specifically, for fault localization, we utilized the spectrum-based (Ochiai) and the eXplainable AI-based (XAI4FL) methods as baselines. For vulnerability detection, we used CodeBERT as a baseline. Across both tasks, the results showed that all LLMs outperformed the non-LLM-based techniques. Our findings in the fault localization revealed that GPT-4o achieved the best results, with a 12.18\% improvement at the Top-1 metric over GPT-3.5, while the smallest LLM in the experiment, Gemma-7B showed a performance decline of 15.2\%. 
In contrast, in the vulnerability detection, Gemma-7B led with a 7.8\% improvement in accuracy over the GPT-3.5 baseline. 
Our analysis suggests that this discrepancy can be attributed to the complexity of the output (i.e., binary vs dynamic output). 
This highlights that while larger LLMs generally tend to perform better on tasks with more complex outputs, it is possible for the smaller models to outperform them on tasks with simpler outputs.

From our comparative analysis, we observed several correct predictions that are unique to the corresponding LLMs. 
To leverage the complementary strengths of these models, we experimented with a voting mechanism to make a final prediction. 
Our findings indicate that combining the results through voting led to an improvement of 13.7\% in the number of faults localized at Top-1 and 11.2\% accuracy in vulnerability detection, compared to the baseline (i.e., GPT-3.5). 

We proposed a validation prompt aimed at further enhancing the results, by presenting output from one LLM to another LLM and prompting it to consider revision of the initial answer (see Section~\ref{sec:approach}). In our experiment, asking GPT-4o to refine its results using additional information from LLaMA-3-70B (GPT-4o $\Leftarrow$ LLaMA-3-70B) led to a 16.2\% improvement in successfully localized faults at Top-1 compared to GPT-3.5, and a minimum 3.6\% improvement over running the LLM individually. Additionally, requiring the LLMs to explain their answers further enhanced results, as demonstrated by GPT-4o $\Leftarrow$ Gemma-7B, which showed improvements of 12\%, 7\%, and 4\% compared to GPT-3.5, GPT-4o, and Gemma-7B, respectively. This highlights the effectiveness of the validation prompt and the importance of explanations to enhance LLM results.

The main contributions of our study are as follows:
\begin{itemize}[]
\item Conducting an evaluation of six LLMs on SQA tasks and found that smaller LLM (Gemma-7B) can outperform larger LLM (LLaMA-3-70B) in the task that required simpler output.
    \item Demonstrating that combining different LLMs through a voting mechanism can enhance the results of both SQA tasks.
    \item Introducing a validation prompt to refine one LLM result using another, demonstrating that a two-LLM combination outperforms a six-LLM ensemble using a voting mechanism.
\end{itemize}

\section{Background and Related Work}
\label{sec:background}

\subsection{Large Language Models (LLMs)}
Large Language Models (LLMs) are trained on extensive text data using advanced deep learning architectures like the Transformer architecture~\cite{vaswani2017attention}. LLMs have showcased remarkable effectiveness across various natural language processing tasks~\cite{kenton2019bert}. LLM also has been used in automated software engineering, such as incident management of cloud service~\cite{ahmed2023recommending}, automated program repair~\cite{xia2023keep, liu2023refining}, code generation~\cite{feng2023investigating}, vulnerability detection~\cite{fu2023chatgpt, zhang2024promptenhanced, zhou2024large}, and fault localization~\cite{wu2023large,kang2023preliminary,widyasari2024demystifying}.

There are various examples of LLMs, among which one prominent model is ChatGPT, developed by OpenAI. ChatGPT is built upon the GPT-3 architecture~\cite{brown2020language} and has been refined through reinforcement learning from human feedback (RLHF). Recently, OpenAI released GPT-4o, which is claimed to outperform GPT-4 in text evaluation~\cite{OpenAI_Models}. 
This new model is available alongside the free version of ChatGPT (GPT-3.5) and the premium version (GPT-4). 
Other significant LLMs include LLaMA 3 developed by Meta. There are two versions of LLaMA-3, one with 8 billion and another with 70 billion parameters~\cite{Meta_LLaMA3}. Another example of LLMs is Mixtral, created by Mistral AI. It adopts a Mixtures of Experts (MoE) architecture, consisting of eight distinct models, each equipped with seven billion parameters~\cite{Mistral_Mixtral}. Google has also made significant contributions to the field with LLMs like Gemini (closed model) and Gemma (open model). Gemma is a recent lightweight state-of-the-art open model that is built following the Gemini models~\cite{Google_Gemma}. In our study, we focus on Mixtral-8x7B, LLaMA-3-70B, LLaMA-3-8B, and Gemma-7B due to their popularity and their availability as open-source models. We also run the closed model GPT-4o in addition to the GPT-3.5 as this model claims to outperform GPT-4 with faster inference. These LLMs offer a diverse range of capabilities and have gained recognition within the community (cf. Section~\ref{sec:llm_setting}).

\subsection{Software Quality Assurance Tasks}
Software Quality Assurance (SQA) is a critical part of the software development life cycle aimed at ensuring that software products meet specified quality standards and functional requirements~\cite{runeson1998software,maxim2016introduction}. The necessity for high-quality, reliable software is critical, as its failure can result in substantial harm and financial loss. For instance in the 2022 report, poor software quality in the United States has grown to at least \$2.41 trillion~\cite{krasner2022cost}. 
SQA encompasses a broad range of tasks designed to ensure software quality. In this study, we focused on two integral examples of SQA tasks that have a direct impact on software reliability and security~\cite{zou2019empirical}.

\subsubsection{Fault Localization} Fault localization is a significant aspect of SQA, it is a task to pinpoint the specific locations in the code that are likely to cause software failures~\cite{yoo2013fault}. This process is essential for efficient debugging and fixing of software, directly impacting the reliability and performance of software products. Over the years, researchers have proposed automatic fault localization techniques to help developers save time in this process. The proposed techniques for fault localization include a variety of methods such as machine-learning-based approaches~\cite{ sohn2017fluccs}, static analysis-based approaches~\cite{neelofar2017improving, feyzi2018fpa}, spectrum-based approaches~\cite{wong2013dstar, abreu2009spectrum, naish2011model}, and so on.  
In this study, we focus on line-level granularity fault localization as it offers the finest resolution and has been widely utilized in previous research~\cite{pearson2016evaluating,wu2023large,kang2023preliminary}.

\subsubsection{Vulnerability Detection.} Vulnerability detection is a critical component of SQA, focusing on identifying whether there are security weaknesses that could be exploited by attackers. This is crucial for preventing potential security breaches and ensuring that software systems remain secure~\cite{li-2022-uchecker}. Vulnerability detection is a highly time-consuming task that requires the reviewers to be knowledgeable about potential security issues~\cite{mcgraw2008automated}. Due to this, many researchers proposed methods to detect the vulnerabilities automatically. There are static analysis-based approaches~\cite{kaur2020comparative, shahriar2012mitigating}, dynamic analysis-based approaches~\cite{aggarwal2006integrating, amin2019androshield}, and so on. The output of vulnerability detection is a binary, vulnerable vs non-vulnerable~\cite{zheng2020impact, grieco2016toward}. 

\subsection{LLM and SQA Tasks}
\subsubsection{LLM and Fault Localization} 
There are several recent studies that utilize LLMs for fault localization. For example, a study by Wu et al.~\cite{wu2023large} provides an evaluation of fault localization using GPT-3.5 and GPT-4. Kang et al~\cite{kang2023preliminary} proposed AutoFL that utilizes ChatGPT to provide fault localization results with test coverage as additional input. Widyasari et al.~\cite{widyasari2024demystifying} proposed FuseFL that utilizes several combinations of information related to the faults (i.e., spectrum-based fault localization results, test case assertion and error, and the code description) to enhance the ChatGPT results. 
There is also a study by Jiang et al.~\cite{jiang2024evaluating} that runs the preliminary evaluation of fault localization and automated program repair using prompts with additional information on the error (i.e., stack trace) and assertion from the test in GPT-3.5 and comparing it with two other closed-source models which are ERNIE, Bot 3.5, and IFlytek Spark 2.0. Furthermore, a recent study by Yang et al.~\cite{yang2024large} built LLMAO an LLM-based fault localization
approach to localize the faulty lines without the need to input any test coverage information.

\subsubsection{LLM and Vulnerability Detection}
Study by Cheskov et al.~\cite{cheshkov2023evaluation} utilize GPT-3.5 for the vulnerability detection and classification of five CWE-IDs. 
A previous study by Fu et al.~\cite{fu2023chatgpt} utilizes GPT-3.5 and GPT-4 to detect, classify, and revise the vulnerability in the code. Tamberh and Bahsi~\cite{tamberg2024harnessing} analyse the vulnerability detection with new prompting techniques such as tree
of thoughts (ToT) and self-consistency using closed-source LLM: GPT-4 and Claude 3 Opus.
A study by Zhang et al.~\cite{zhang2024prompt} also explores the use of ChatGPT in vulnerability detection through structural and sequential auxiliary information. 
There is also a recent work~\cite{sun2024gptscan} that proposed GPTScan, which combines GPT with static analysis for vulnerability detection. Recent work by Zhou et al.~\cite{zhou2024large} integrates knowledge from the CWE system and similar samples to enhance the prompt results using GPT-3.5 and GPT-4.  

\vspace{0.1cm}
In previous studies using LLMs for SQA tasks, ChatGPT is often used as the engine. However, how do these results compare with other recent open-source LLMs? In this study, we explored various LLMs and proposed a method to combine their results.

\section{Cross-Validation Technique}
\label{sec:approach}
\begin{figure*}
\centering
\includegraphics[width=0.98\textwidth, angle=0]{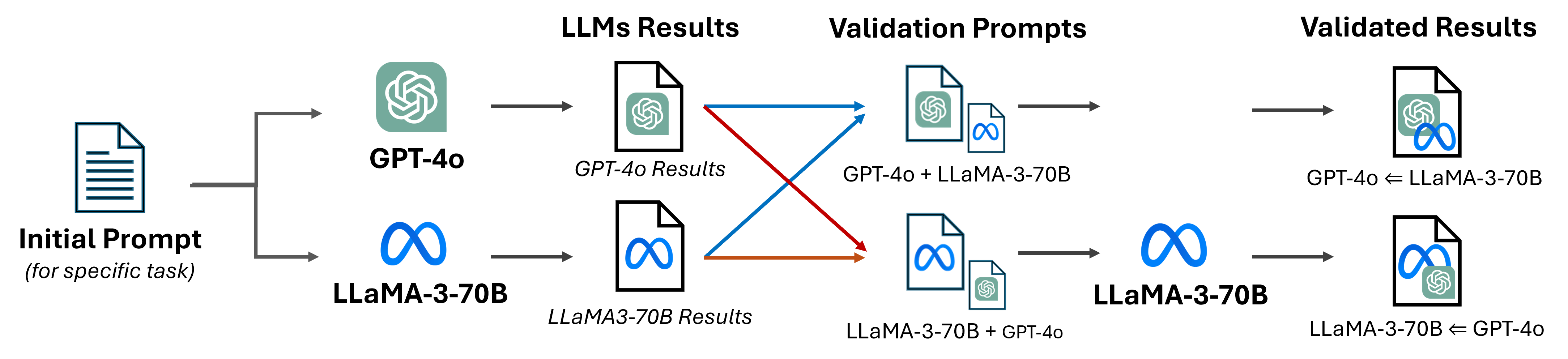}
\caption{Illustration of the proposed approach on cross-validating GPT-4o and LLaMA-3-70B}
\label{fig:architecture}
\end{figure*}

In this section, we describe our approach for validating results obtained from one LLM using another. 
The illustration of the proposed approach in this paper is highlighted in Figure~\ref{fig:architecture}. Our approach leverages the strengths of various LLMs to iteratively enhance the accuracy of the final results. 
The cross-validation approach required us to do multiple rounds of prompting, we first utilize \textit{initial prompts} to obtain base results from the LLMs. These \textit{initial prompts} refer to the prompts that we used to derive results for the specific tasks (i.e., fault localization and vulnerability detection). Then, we validate the results from one LLM with those from another using our proposed \textit{validation prompt}. For instance, results from GPT-4o can be validated using those from LLaMA-3-70B. This iterative process allows for refinement by incorporating insights from different LLMs. Details regarding the \textit{initial prompts} used in this study and the proposed \textit{validation prompts} are provided in the next subsection.

\subsection{Initial Prompts}\label{sec:initial_prompts}

\subsubsection{Fault Localization}
For the \textit{initial prompt} in fault localization (Table~\ref{table:initial_prompts}), we utilized the prompt from the recent previous study~\cite{widyasari2024demystifying}. The study introduces FuseFL that uses a prompt that is a combination of several pieces of information: suspicious line ranking obtained from statement-based fault localization (SBFL) results, input and output obtained from the test cases, and the code description (i.e., an explanation of the intended purpose of the provided code). For the SBFL result, we used the Top-5 results from Ochiai~\cite{abreu2006evaluation} following the previous study~\cite{widyasari2024demystifying}. It also utilizes zero-shot chain-of-thought~\cite{kojima2022large} prompting, so the LLM needs to provide reasoning on why the location is considered faulty.

\begin{table}[]
\caption{Initial prompts used in the study}
\label{table:initial_prompts}
\small
\begin{tabularx}{0.97\textwidth}{|C|}
\hline
\multicolumn{1}{|c|}{\cellcolor[HTML]{D0D0D0}{ \textbf{Fault Localization}}}   \\ \hline
\textbf{Faulty Code}: [Faulty Code] \\

\textbf{Code Description}: [Code Description] \\

\textbf{Test Results}: The code is producing incorrect results. \\

  - Running the function as follows ‘[Input]‘ generate an ‘[Error name]‘ in line [i] ‘[Code content]‘. \\

  - Running the function as follows ‘[Input]‘ yields‘[Output]‘ instead of the expected ‘[Expected Output]‘. \\

\textbf{Spectrum-based Fault Localization Techniques Results}: \\

  1. Line [i] ‘[Code]‘, [Tech. name] score: [Score] ... \\

Analyze the provided code and utilize the code description, test results, and SBFL techniques results to help in identifying potentially faulty lines. Provide the results following this JSON template: 
\{"faultLoc": [ \{ "faultyLine": (line number of the suspicious code), 
"code": (displaying the actual code), 
"explanation": (\textbf{step-by-step reasoning} on why this location is considered potentially faulty) \}, ... ] \}. Ensure that the objects in the "faultLoc" array are sorted in descending order of suspicion. \\
\hline
\multicolumn{1}{|c|}{\cellcolor[HTML]{D0D0D0}{ \textbf{Vulnerability Detection}}}  \\ \hline
Here are \textbf{sampled examples from the training data}.\\ Example1: [Example Code], Label1: this code is [Vulnerable/Non-vulnerable].  ...\\ Here are the \textbf{most similar codes from the training data}.\\ Example1: [Example Code], Label1: this code is [Vulnerable/Non-vulnerable]  ... \\
Now you need to identify whether a method contains a vulnerability or not. If has any potential vulnerability, output: `this code is vulnerable’. Otherwise, output:  `this code is non-vulnerable’. \\The code is {[}Code{]}
\\
\hline
\end{tabularx}
\end{table}

\subsubsection{Vulnerability Detection}
For the \textit{initial prompt} in the vulnerability detection task (Table~\ref{table:initial_prompts}),  we use prompt A54 from a recent study by Zhou et al.~\cite{zhou2022large}, which demonstrated the best performance in their study.  
The prompt instructed the LLMs to analyze the input code and determine whether "the code is vulnerable" or "the code is non-vulnerable." Then, for the additional information, it includes two sets of examples: the first set comprises three randomly selected code examples from the training data, and the second set consists of three most similar code examples from the training data, matched to the corresponding code. The similar codes are identified by calculating the cosine similarity of the semantic vectors, which are generated using CodeBERT from the code samples.

\subsection{Validation Prompt}

\begin{figure}[hbtp]
    \includegraphics[scale=0.3]{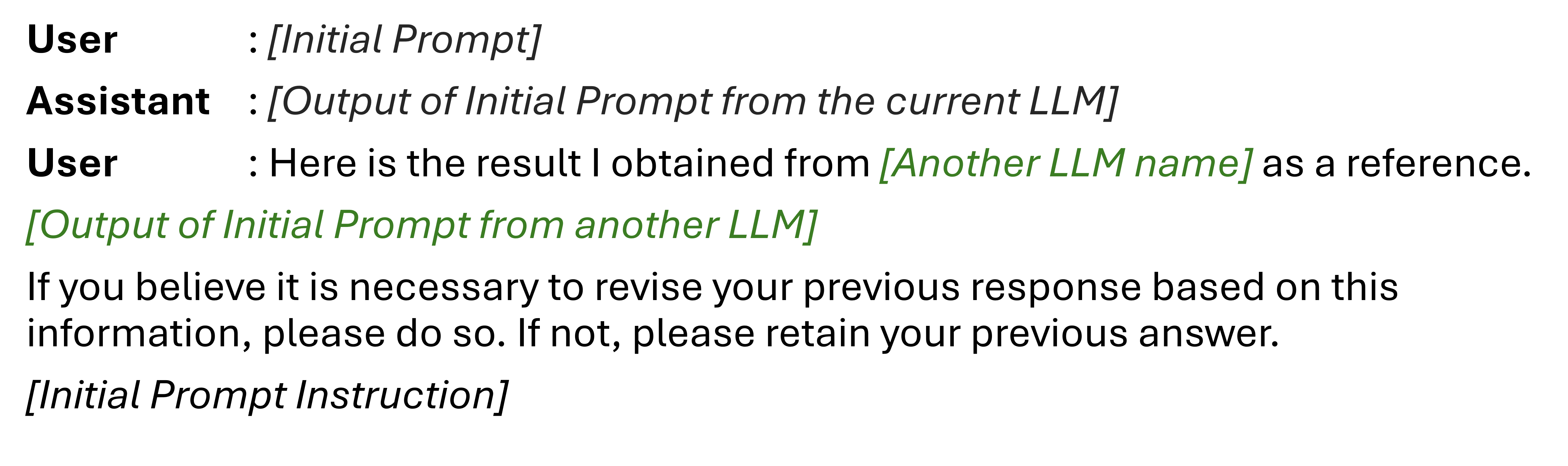}\\
    \caption{Validation prompt}
    \label{fig:validation_prompt}
\end{figure}

The validation prompt we proposed is illustrated in Figure~\ref{fig:validation_prompt}. This validation prompt uses the results from another LLM to challenge the current LLM to validate or revise its previous response based on insights from another LLM.  
In cases where the LLM lacks inherent support for tracking historical interactions, we address this limitation by incorporating the previous response directly into the validation prompt. This is structured into separate ``User'' and ``Assistant'' sections, which helps to systematically integrate historical interaction (i.e., the initial prompt that the user used and the output from the corresponding LLM) into the validation process. The ``Assistant'' section in the prompt refers to the target LLM. 

Furthermore, we introduced the \textit{ensure} instruction in the validation prompt, which reiterates key aspects of the previously used initial prompt. For instance, in the validation prompt for the fault localization task, we include the specific instruction taken from the initial prompt: ``\textit{Ensure that the objects in the `faultLoc' array are sorted in descending order of suspicion.''} This addition helps ensure that the LLMs comprehend and remember the crucial requirement of prioritizing the suspicious locations in descending order for the fault localization task. Similarly, for the vulnerability detection task, we reemphasize the core instruction from the initial prompt, reminding the LLMs to check whether the code contains potential vulnerabilities: \textit{``If it has any potential vulnerability, output: `this code is vulnerable'. Otherwise, output: `this code is non-vulnerable’.''} To make sure that the additional \textit{ensure} component in the prompt improves the results (i.e., prompt engineering). We conducted additional experiments in two best-performing LLMs with and without \textit{ensure} component. 
Our evaluations highlighted in Tables~\ref{tab:prompt_engineering_FL} and~\ref{tab:prompt_engineering_VD} indicate that including \textit{ensure} component yields better overall results. Based on these findings, we used the \textit{ensure} component in subsequent experiments.

\begin{table}[]
\caption{Cross-Validation results for fault localization with and without \textit{ensure} component}
\label{tab:prompt_engineering_FL}
\centering
\begin{tabular}{|l|r|r|r|}
\hline
\textbf{Fault Localization}                   & \textbf{Top-1} & \textbf{Top-2} & \textbf{Top-3} \\ \hline
LLaMA-3-70B$\Leftarrow$GPT-4o  & 221                                & 260                                & 280                                \\ 
LLaMA-3-70B$\Leftarrow$GPT-4o + \textit{Ensure}   & \textbf{224}                       & \textbf{265}                       & \textbf{284}                       \\ \hline
GPT-4o$\Leftarrow$LLaMA-3-70B  & 226                                & 270                                & \textbf{288}                       \\ 
GPT-4o$\Leftarrow$LLaMA-3-70B + \textit{Ensure}   & \textbf{229}                       & \textbf{270}                       & 283                               \\ \hline
\end{tabular}
\end{table}

\begin{table}[]
\caption{Cross-Validation results for vulnerability detection with and without \textit{ensure} component}
\label{tab:prompt_engineering_VD}
\centering
\begin{tabular}{|l|r|r|r|r|r|}
\hline
\textbf{Vulnerability Detection}                & \textbf{Acc} & \textbf{Prec} & \textbf{Rec} & \textbf{F1} & \textbf{F0.5} \\ 
\hline
Gemma-7B$\Leftarrow$GPT-4o & 67.4                                  & 70.6                                   & 59.6                                & 64.6                            & 68.0                              \\
Gemma-7B$\Leftarrow$GPT-4o + \textit{Ensure}  & \textbf{67.9}                         & \textbf{70.7}                          & \textbf{61.1}                       & \textbf{65.6}                   & \textbf{68.5}                     \\ 
\hline
GPT-4o$\Leftarrow$Gemma-7B  & \textbf{67.9}                         & 68.3                                   & \textbf{66.8}                       & \textbf{67.5}                   & 68.0                              \\
GPT-4o$\Leftarrow$Gemma-7B + \textit{Ensure}   & \textbf{67.6}                         & \textbf{71.0}                          & 59.6                                & 64.8                            & \textbf{68.4}              \\ \hline      
\end{tabular}
\end{table}

We hypothesize that this validation prompt can lead to further improvements in the results. By prompting the current LLM to reevaluate its initial answer and potentially refine it based on insights from another LLM, we introduce an additional layer of refinement. Additionally, integrating perspectives from different LLMs provides a diverse range of insights, potentially enhancing the final results. This iterative validation process encourages a more thorough examination of the code and supports collaboration between different LLMs. 

\section{Research Questions and Experiment Settings}
\label{sec:data}
\subsection{Research Questions}
\subsubsection{RQ1} \textit{How effective are various Large Language Models (LLMs) in performing fault localization and vulnerability detection tasks?}

In recent studies on fault localization and vulnerability detection using LLMs, ChatGPT has predominantly served as the engine for these approaches.  The results from using ChatGPT for these tasks have shown improvements compared to baseline methods. However, the performance of these tasks across various LLMs remains largely unexplored. Our study aims to address this gap by comparing fault localization and vulnerability detection performance using different LLMs. We employ several LLMs (Section~\ref{sec:llm_setting}), including GPT-4o, LLaMA-3-70B, LLaMA-3-8B, Mixtral-8x7B, and Gemma-7B, in comparison with GPT-3.5 as the baseline. 
For the prompts, we utilize those from recent studies~\cite{widyasari2024demystifying, zhou2024large}, which have demonstrated improvements over state-of-the-art baselines (Section~\ref{sec:initial_prompts}). 
Furthermore, we also utilized the state-of-the-art non-LLM-based techniques for both task as a baseline. For fault localization, we utilized the SBFL technique, Ochiai~\cite{abreu2006evaluation}, and the eXplainable AI-based technique, XAI4FL~\cite{widyasari2022xai4fl}, which have been used in recent previous studies~\cite{widyasari2024demystifying, zeng2022fault, yang2021evaluating}. For vulnerability detection, we also compared LLMs results with CodeBERT, a state-of-the-art method according to a recent comprehensive empirical study~\cite{steenhoek2023empirical}. 
Through this comparative analysis, we aim to provide insights into the effectiveness of different LLMs for software quality assurance tasks.

\subsubsection{RQ2} \textit{How effective is the combination of multiple LLMs results using voting for the selected Software Quality Assurance (SQA) tasks?}\label{sec:rq2_method}

For this research question, our goal is to analyze the results of combining different LLMs results using a voting mechanism, specifically majority voting. In the fault localization task, the Top-1 location is determined based on the most frequently voted top-1 locations across different LLMs. For instance, if GPT-4o, LLaMA-3-70B, Gemma-7B, and Mixtral-8x7B identify lines 2, 3, 2, and 2 as the most suspicious location, the consolidated Top-1 prediction would be line 2. To streamline the process, the final fault locations for Top-1, Top-2, and Top-3 are calculated independently. For instance, if line 3 is identified as the Top-1 by LLaMA-3-70B but is not selected as the final Top-1 location, it will not be considered in the subsequent rank calculations (Top-2 and Top-3). In the event of a tie, where two or more candidate locations receive an equal number of votes, we prioritize the location identified by GPT-4o, as it is the most resource-intensive model in our study.

Similarly, for vulnerability detection, we apply majority voting to determine if the code is vulnerable. For example, if the predictions from various LLMs are vulnerable, vulnerable, not vulnerable, and vulnerable, the aggregated result would classify the code as vulnerable. 
In case of a tie, following the fault localization, we utilized prediction from GPT-4o. 
This research question is crucial as it evaluates the potential of combining various LLM results to enhance the accuracy of fault localization and vulnerability detection.

\subsubsection{RQ3} \textit{How effective is the proposed LLMs cross-validation technique in the fault localization and vulnerability detection tasks?}

We seek to determine whether we can achieve improved results by leveraging our proposed cross-validation technique in both tasks. Our proposed technique involves utilizing one LLM to validate the findings of another LLM (Section~\ref{sec:approach}). For the experiment, we rank the results the results from RQ1 and divide them into three groups: low-ranked, medium-ranked, and high-ranked performance. We then perform cross-validation within these groups. For instance, if the performance ranking of LLMs is GPT-4o, LLaMA-3-70B, Mixtral-8x7B, GPT-3.5, LLaMA-3-8B, and Gemma-7B, we will cross-validate between (1) GPT-4o and LLaMA-3-70B; (2) Mixtral-8x7B and GPT-3.5; (3) LLaMA-3-8B and Gemma-7B. We are adopting this approach to make the results more concise. By grouping the LLMs, we can ensure that the models within each group do not have significant performance differences, thereby optimizing the cross-validation process. This research question is significant as it explores a method to enhance fault localization and vulnerability detection accuracy by leveraging complementary insights from different LLMs.

\subsubsection{RQ4} \textit{How does the explanation provided by LLMs impact the effectiveness of cross-validation techniques}

Our objective is to determine whether the explanations given by LLMs can enhance the final results obtained from cross-validation techniques. In the initial prompt used for fault localization~\cite{widyasari2022xai4fl}, it used Chain-of-Thought (CoT) prompting~\cite{kojima2022large}, which requires the LLM to provide step-by-step reasoning for its answers. In contrast, in the vulnerability detection prompt taken from a previous study~\cite{zhou2024large}, the LLM was instructed to simply indicate whether a code snippet was vulnerable or not, without providing reasoning. 
To investigate the impact of explanations, we also instructed the LLMs in the vulnerability detection task to provide \textit{``step-by-step reasoning on why the code is potentially vulnerable or non-vulnerable.''} 
By comparing scenarios where LLMs provide reasoning against those where they do not, we can better understand the contribution of explanations to enhancing and refining answers in SQA tasks.

\subsection{Dataset}\label{sec:dataset}
\textit{Fault Localization.} We utilize the dataset from the previous study that introduces FuseFL~\cite{widyasari2024demystifying}, which we used as the initial prompt in our study. The FuseFL dataset is derived from the faulty code in the \textit{Refactory} dataset~\cite{hu2019re}, which includes submissions from an introductory programming course. These submissions, typically from novice developers, often contain errors and illustrate scenarios where feedback on code is vital. The dataset consists of 324 faulty Python code samples, encompassing a total of 600 faulty lines. More than half of the faulty code (57\%) in the dataset has multiple faults. This is inline with the previous study~\cite{digiuseppe2011influence}, which highlighted that the majority of real-world software often contains multiple faults. The dataset captures a wide range of issues, which are categorized into two main types: RuntimeFaults (41.4\%), where the code cannot execute to completion, and OutputFaults (58.6\%), where the code's output deviates from the developer's expected result.
\\[1pt]
\noindent\textit{Vulnerability Detection.} We use the dataset provided in the previous study by Xin et al.~\cite{zhou2024large}, from which we also adopt the proposed prompts. It originated from a vulnerability-fixing commit dataset collected by Pan et al.~\cite{pan2023fine}. This dataset includes a total of 386 methods in the test set, with 193 of these methods being vulnerable functions and the remaining 193 being non-vulnerable functions. The code samples are sourced from 20 different software repositories implemented in C/C++.This dataset featured 9 distinct Common Weakness Enumerations (CWEs), of which 6 are among the top 25 most dangerous CWE.\footnote{\url{https://cwe.mitre.org/top25/archive/2023/2023_top25_list.html}}

\subsection{Evaluation Metrics}\label{sec:metrics}
\textit{Fault Localization.} To evaluate the effectiveness of the fault localization technique in identifying faults, we employ Top-K metrics that have been used in many previous studies~\cite{kochhar2016practitioners,pearson2016evaluating,widyasari2022xai4fl, widyasari2024demystifying, jiang2024evaluating,yang2024large}. The Top-K score indicates how often the fault localization technique successfully identifies faults within the top K positions of its rankings.
Given that our dataset includes multiple fault lines per sample, we employ a best-case scenario approach for our evaluation~\cite{pearson2016evaluating,widyasari2022xai4fl}. In this scenario, the fault localization is considered successful if the technique identifies any of the fault lines within the top K most suspicious lines. We use K values of 1, 2, and 3, with Top-1 serving as the main metric~\cite{widyasari2022xai4fl,widyasari2024demystifying,jiang2024evaluating}. The Top-1 metric is particularly important in automated program repair, as it often assumes perfect fault localization (i.e., identified in the Top-1 position)~\cite{chen2019sequencer,lutellier2020coconut,jiang2021cure}.

\textit{Vulnerability Detection.} For the vulnerability detection task, we use metrics (i.e., Accuracy, Precision, Recall, F1, and F0.5) commonly employed in previous studies~\cite{zhou2024large, zhang2024promptenhanced}. The formulas for these metrics are as follows:
\[
\text{Accuracy} = \frac{TP + TN}{TP + TN + FP + FN}
\quad \quad
\text{Precision} = \frac{TP}{TP + FP}
\]
\[
\text{Recall} = \frac{TP}{TP + FN}
\quad \quad \quad \quad \quad \quad
\text{F1} = 2 \cdot \frac{\text{Precision} \cdot \text{Recall}}{\text{Precision} + \text{Recall}}
\]
\[
\text{F0.5} = (1 + 0.5^2) \cdot \frac{\text{Precision} \cdot \text{Recall}}{0.5^2 \cdot \text{Precision} + \text{Recall}}
\quad\quad\quad\quad\quad\quad\quad
\]

The F1 metric balances precision and recall equally, while the F0.5 metric places more emphasis on precision. The primary metric for discussing the results is accuracy, following the discussions in previous studies~\cite{zhou2024large, zhang2024promptenhanced}.

\subsection{LLMs Setting}\label{sec:llm_setting}
Our model selection includes a diverse range of available LLMs. ChatGPT, encompassing both GPT-3.5 and GPT-4o, is a closed-source LLM with no publicly available details on its technical specifications. 
In contrast, the other four LLMs are publicly available but vary in their sizes and structure. LLaMA-3-70B is a single model with a substantial size of 70 billion parameters. Mixtral employs a Mixture of Experts (MoE) approach, utilizing a combination of eight models, each with 7 billion parameters. Meanwhile, Gemma-7B and LLaMA-3-8B are single lightweight models with 7 and 8 billion parameters, respectively. 
These models have been popularly used across various domains~\cite{zhu2024openai, mo2024fine, chopra2024house, yoon2024bigger, adams2024llama}. The following section provides further details on the LLMs utilized in this study.  

\textit{ChatGPT~\cite{OpenAI_Models}} is a closed LLM model developed by OpenAI. It is one of the most popular LLMs. 
In our study, we use the GPT-3.5 model, specifically {\tt gpt-3.5-turbo-0125}, which provides a context window of 16,385 tokens. Additionally, we also utilize a newer model from OpenAI, GPT-4o, which is reported to exceed the performance of GPT-4 based on their benchmark results~\cite{gpt4o_evaluation}. GPT-4o is also faster and has greater accessibility compared to the GPT-4 model. Specifically, we use {\tt gpt-4o-2024-05-13}, that have a context window of 128,000 tokens. 

\textit{LLaMA-3~\cite{Meta_LLaMA3}} is a new publicly available LLM developed by Meta. 
LLaMA-3 has been trained on a substantial corpus of 15 trillion tokens, which is a significant increase from its predecessor's (i.e., LLaMA 2) 2 trillion tokens. LLaMA 3 employs a tokenizer with a vocabulary of 128K tokens that can encode language more efficiently. For our study, we utilize two models from LLaMA-3, {\tt llama-3-70b-8192} model and {\tt llama-3-8b-8192} model which have 70B and 8B parameters, respectively. These models have a context window of 8,192 tokens.

\textit{Mixtral~\cite{Mistral_Mixtral}} developed by Mistral AI, is a Mixtures of Experts (MoE) LLM that consists of 8 distinct models, each with 7 billion parameters. In the MoE architecture, the feedforward block of the network will choose two of these models to process and combine the output tokens. The Mixtral model is pre-trained on data extracted from the open web. In our study, we use the {\tt Mixtral-8x7B-Instruct\-v0.1} model, the instruction-tuned version of the Mixtral model that allows for a context length of 32,768 tokens.

\textit{Gemma~\cite{Google_Gemma}} is a series of publicly available LLMs developed by Google. Compared to the other LLMs that we use, Gemma is considered a lightweight LLM, as its largest model only has 7 billion parameters. The 1.1 version of Gemma models is trained using the RLHF method on up to 6 trillion tokens of text data consisting of web documents, code, and mathematics data. For our study, we utilize the {\tt gemma-7b-it} model, the latest and largest available instruct version of the Gemma LLM with a context window of 8,192 tokens.

\section{Results}
\label{sec:results}
\subsection{RQ1: LLMs performance in SQA Tasks}\label{sec:rq1_results}
The results of fault localization using different LLMs (i.e., GPT-3.5, GPT-4o, LLaMA-3-8B, LLaMA-3-70B, Gemma-7B, and Mixtral-8x7B) and non-LLM-based techniques (i.e., Ochiai and XAI4FL) are presented in Table~\ref{tab:RQ1_topk}. We found that all LLMs used in this study outperformed Ochiai and XAI4FL in the Top-1 metric, with improvements ranging from  12.08\% to 88.89\%. 
These results highlight the capability of LLMs to enhance fault localization in this dataset. 
Among the LLMs, GPT-4o achieved the highest number of successfully localized faults at the Top-1 position, followed by LLaMA-3-70B, Mixtral-8x7B, GPT-3.5, LLaMA-3-8B, and Gemma-7B, in descending order. The difference between GPT-4o and LLaMA-3-70B was minimal, with GPT-4o localizing just one more fault. GPT-4o showed a 12.18\% improvement over GPT-3.5 at Top-1. Meanwhile, LLaMA-3-8B and Gemma-7B showed a notable decline in Top-1 performance of 12.56\% and 15.22\%, respectively, when compared to GPT-3.5.

The lower performance of Gemma-7B and LLaMA-3-8B may be attributed to its smaller model size compared to other models. Gemma-7B only has 7 billion parameters compared to the substantially larger models of LLaMA-3-70B and Mixtral-8x7B, which have 70 billion and a total of 56 billion parameters (8x7), respectively. 

\begin{table}[h]
\caption{Fault localization results}
\label{tab:RQ1_topk}
\centering
\begin{threeparttable}
\centering
\begin{tabular}{|l|l|l|l|l|}
\hline
\textbf{Technique}     & \textbf{Top-1} & \textbf{Top-2} & \textbf{Top-3} &  \textbf{$\Delta$ GPT-3.5 \tnote{\S}} \\ \hline
Ochiai & 117  & 208 & 247                   & -68.38\%                         \\ 
XAI4FL & \textbf{149}  & 216  & 248                    & -32.21\%                         \\ \hline
GPT-3.5 & 197  & 240 & 260                    & -                         \\ 
LLaMA-3-8B            & 172 & 212 &229                      & -12.69\%                   \\ 
Gemma-7B            & 167 & 207 & 229                      & -15.22\%                  \\ 
Mixtral-8x7B          & 199 &      235 & 252                & +1.01\%                   \\ 
LLaMA-3-70B         & 220    & \textbf{267} & \textbf{283}                  & +11.68\%                   \\ 
GPT-4o         & \textbf{221}  & 259 & 278                    & +12.18\%                   \\ \hline
Comb. LLMs      & 214      & 252 & 270                  & +8.62\%                   \\ 
Comb. (w/o Gemma)     & \textbf{224}      & 256 & 276                  & +13.71\%                   \\ \hline
\end{tabular}
\begin{tablenotes}
        \footnotesize
        \item[\S] Relative improvement of the Top-1 result compared to the GPT-3.5 that being used in the previous studies~\cite{widyasari2020bugsinpy,jiang2024evaluating}.       
    \end{tablenotes}
\end{threeparttable}
\end{table}

The dataset can be categorized into two main fault types: RuntimeFaults, where the code cannot execute to completion, and OutputFaults, where the code's output deviates from the expected result. We found that smaller models (e.g., LLaMA-3-8B and Gemma-7B) perform poorly in localizing OutputFaults type. For instance, when comparing Gemma-7B, the smallest and least performant model, with GPT-4o, the highest performer, we observed a 10\% difference in RuntimeFaults and a much larger 65\% difference in OutputFaults. This underscores that while smaller models may be adequate for detecting RuntimeFaults, they struggle considerably with OutputFaults in comparison to larger models. Unlike RuntimeFaults, which have clear indicators like halted execution, 
OutputFaults require a deeper understanding of the code's intended logic and the ability to compare the actual output against the expected output. This process involves multiple layers of reasoning and often requires the model to \textit{``simulate''} the code to detect where the logic deviates. Smaller models 
may struggle with this level of reasoning and inference.

\begin{figure}
\centering
\includegraphics[scale=0.55]{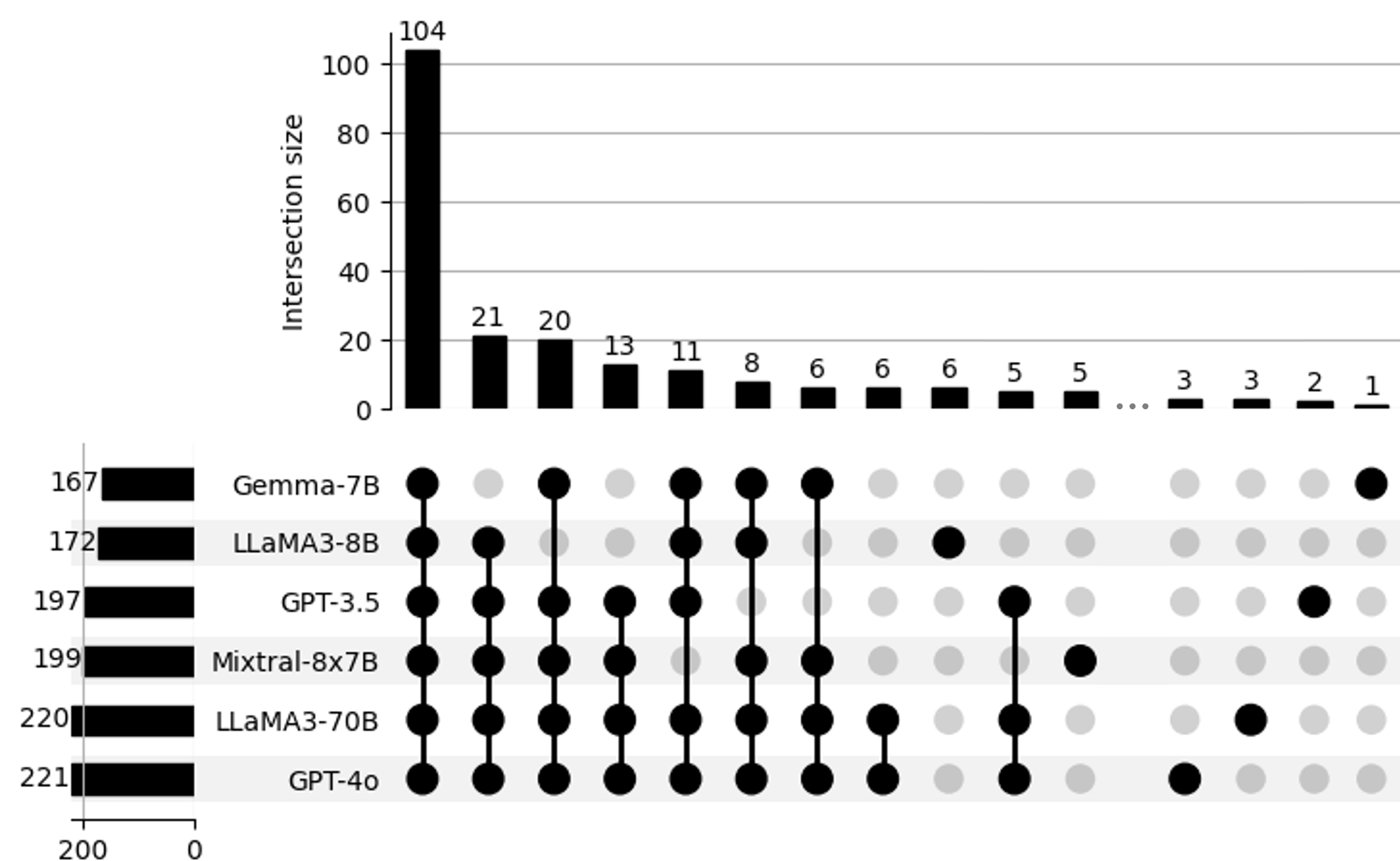}
\caption{Intersection diagram showing successfully localized faults at Top-1\protect\footnotemark
}
\label{fig:venn_rq1}
\end{figure}

Figure~\ref{fig:venn_rq1} shows the intersection diagram of the faults that are successfully localized at Top-1 by all the LLMs in our study. We discovered that several faults are uniquely localized by specific LLMs. 
Interestingly, LLaMA-3-8B, which ranks second to last in terms of overall Top-1 fault localization accuracy, has the highest number of faults uniquely localized by it. Meanwhile, Gemma-7B which has the lowest performance, also has the lowest percentage of uniquely localized faults. Additionally, we identified multiple sets of LLMs that successfully localized the same faults. For example, Mixtral-8x7B, GPT-4o, GPT-3.5, LLaMA-3-8B, and LLaMA-3-70B jointly localized 21 faults at the Top-1 position. These shared sets mostly localized more LogicalFaults than RuntimeFaults, especially in sets excluding Gemma-7B. For instance, the set consisting of GPT-3.5, GPT-4o, Mixtral-8x7B, and LLaMA-3-70B localized the same 13 faults, 12 of which were LogicalFaults.

\footnotetext{\label{footnote:1}For better visual, this diagram includes only sets with values greater than 4, as well as single LLM results (e.g., faults localized only by GPT-4o).  
Dark circles denote sets that are part of an intersection.}

For vulnerability detection, the results are presented in Table~\ref{tab:RQ1_vulnerability}. All LLMs in this study achieved higher accuracy in detecting vulnerable methods compared to CodeBERT, with improvements ranging from 1\% to 12.1\%.  
Gemma-7B and GPT-4o achieved the highest accuracy, outperforming the GPT-3.5 baseline by 7.8\% and 7.4\%, respectively. Mixtral-8x7B and LLaMA-3-70B showed slight increases in accuracy over GPT-3.5, by 0.8\% and 1.6\%. 
However, we see a decrease in performance for LLaMA-3-8B to the GPT-3.5 baseline, with reductions of 2.9\% respectively. Although Gemma-7B outperformed GPT-4o in overall accuracy, GPT-4o was more effective in correctly identifying vulnerable cases, suggesting a higher tendency to predict vulnerabilities, which resulted in lower accuracy but higher recall. 

In contrast to the fault localization results, in vulnerability detection, the smaller model Gemma-7B outperforms the larger models such as LLaMA-3-70B and Mixtral-8x7B. 
This variation may be attributed to the complexity of the output tasks. 
The output choices for vulnerability detection are binary, whereas fault localization involves multiple potential outputs depending on the code length and the ranking of suspicions. 
This underscores that while larger LLMs typically excel in tasks with more complex outputs, smaller models can sometimes outperform them on tasks with simpler outputs. 
These findings emphasize the importance of evaluating multiple LLMs across various tasks, as their performance can vary.

Relating the CWE type with the LLM results, we found that all LLMs, except Gemma-7B, performed better on more general CWEs. For instance, CWE-119 is a broad category encompassing vulnerabilities related to improper restriction within allocated memory bounds. This CWE-119 has several children including CWE-125 (out-of-bounds read) and CWE-787 (out-of-bounds write). We observed that LLMs performed better on CWE-119 than on its children. For example, LLaMA-3-70B correctly predicted 80\% cases in CWE-119 but only 30\% in both CWE-125 and CWE-787. This suggests that while general vulnerabilities may be easier for LLMs to identify, cases with more specific CWEs might require specialized models.

\begin{figure}
\centering
\includegraphics[scale=0.5]{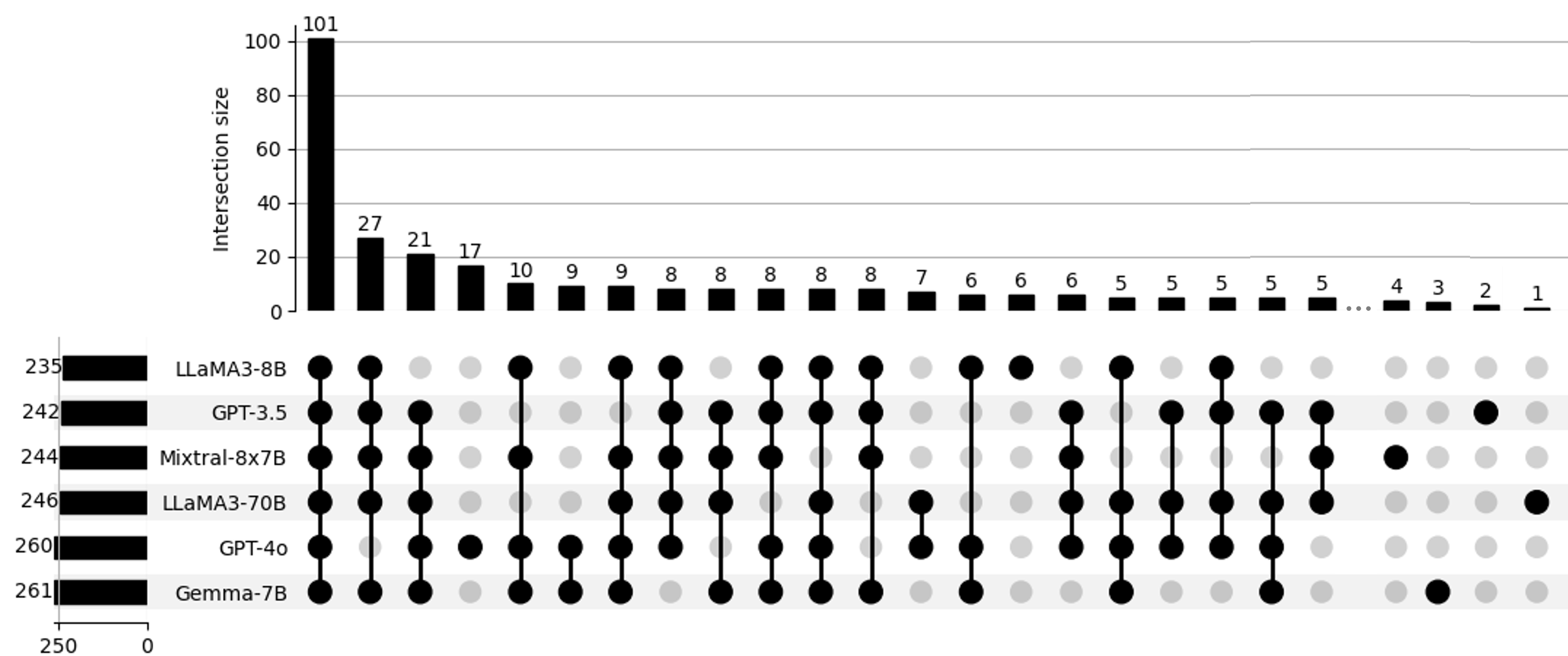}
\caption{Intersection diagram showing correct predictions in vulnerability detection\textsuperscript{\ref{footnote:1}}
}
\label{fig:venn_rq1_vulnerability}
\end{figure}

For further analysis, we also examined the codes that were correctly predicted by the LLMs, as shown in the intersection diagram in Figure~\ref{fig:venn_rq1_vulnerability}. Similar to the fault localization results, we found that different LLMs uniquely predicted certain codes correctly. GPT-4o, which ranked second in accuracy, had the highest number of uniquely correct predictions (17). Interestingly, LLaMA3-8B, despite having the lowest overall accuracy, had the second-highest number of uniquely correct predictions. In contrast, LLaMA-3-70B, with the third-highest accuracy, made only one uniquely correct prediction. We also observed that certain LLMs correctly predicted the vulnerabilities of the same codes. For example, all the LLMs, except GPT-4o, correctly identified the vulnerabilities in the same 27 codes.  These findings, consistent across both vulnerability detection and fault localization, suggest that even lower-performing LLMs can have a unique correct prediction. Additionally, the joint success of certain combinations of LLMs suggests the potential for further improvement by combining outputs from different LLMs. Each model may offer unique insights that can contribute to overall performance enhancement.

\begin{table}[]
\caption{Vulnerability detection results}
\label{tab:RQ1_vulnerability}
\centering
\begin{threeparttable}
    
\begin{tabular}{|l|r|r|r|r|r|r|}
\hline
 \textbf{Technique}                 & \multicolumn{1}{c|}{\textbf{Acc}} & \multicolumn{1}{c|}{\textbf{Prec}} & \multicolumn{1}{c|}{\textbf{Rec}} & \multicolumn{1}{c|}{\textbf{F1}} & \multicolumn{1}{c|}{\textbf{F0.5}} & \multicolumn{1}{l|}{\textbf{$\Delta$GPT-3.5\tnote{\S}}} \\ \hline
 CodeBERT & \textbf{60.3} & 62.3 & 53.3 & 57.3 & 60.1 & 3.9\%\\ \hline
 GPT-3.5                      & 62.7                                   & 76.3                                    & 36.8                                 & 49.7                             & 62.8                               & 0.0\%                                        \\ 
                       LLaMA-3-8B                    & 60.9                                   & 63.5                                    & 51.3                                 & 56.7                             & 60.6                               & -2.9\%                                       \\ 
                       Gemma-7B                     & \textbf{67.6}                          & 71.8                                    & 58.0                                 & 64.2                             & 68.5                     & 7.8\%                                        \\ 
                       Mixtral-8x7B                 & 63.2                                   & 73.4                                    & 41.5                                 & 53.0                             & 63.6                               & 0.8\%                                        \\  
                       LLaMA-3-70B                   & 63.7                                   & 77.9                           & 38.3                                 & 51.4                             & 64.6                               & 1.6\%                                        \\ 
                       GPT-4o                       & 67.4                                   & 66.8                                    & 68.9                        & 67.9                    & 67.2                               & 7.4\%                                        \\ \hline 
                       Comb. LLMs              & \textbf{69.7}                          & 78.8                                    & 53.9                                 & 64.0                             & 72.1                      & 11.2\%                                       \\ 
                       Comb. (w/o L-70B) & 68.4                                   & \textbf{75.2}                           & 54.9                                 & 63.5                             & 70.0                               & 9.1\%                                        \\ \hline

\end{tabular}
\begin{tablenotes}[para]
        \footnotesize
        \item[\S] Percentage increase of the accuracy compared to the GPT-3.5 that being used in the previous studies~\cite{zhou2024large}.
    \end{tablenotes}
\end{threeparttable}
\vspace{-0.2cm}
\end{table}

\begin{tcolorbox}
{\textbf{RQ1 Findings:} 
For fault localization, GPT-4o and LLaMA-3-70B provide the best results and outperform GPT-3.5 by 12.18\% and 11.68\%.
For vulnerability detection, Gemma-7B and GPT-4o outperform GPT-3.5 by 7.8\% and 7.6\%. 
}
\end{tcolorbox}

\subsection{RQ2: Combining LLMs Through  Voting}\label{sec:rq2_results}

The result of combining the fault localization outputs from various LLMs using the voting mechanism is highlighted in Table~\ref{tab:RQ1_topk}. It outperforms Ochiai and XAI4FL by 82.9\% and 43.6\%, respectively. Comparing with the LLMs results, we found that although the combination of all LLMs improved upon GPT-3.5 by 8.62\%, it was still less effective than running LLaMA-3-70B or GPT-4o individually. From the RQ1 analysis, we found that Gemma-7B has the smallest number of faults uniquely localized at Top-1. Gemma-7B also has the largest percentage (62.3\%) of faults that are localized correctly by all the LLMs. These findings suggest that it may not be helpful to include Gemma-7B in the combination. By combining GPT-4o, GPT-3.5, LLaMA-3-8B, Mixtral-8x7B, and LLaMA-3-70B, we achieved the best results with a 13.71\% improvement over GPT-3.5. This combination also outperformed all individual LLMs, with improvements ranging from 1.4\% to 30.3\%.

The results of combining the vulnerability detection outputs from various LLMs are highlighted in Table~\ref{tab:RQ1_topk}. We found that combining all LLMs provided the best accuracy, with an improvement of 15.6\% and 11.2\% over CodeBERT and GPT-3.5 baseline, respectively. This result also improved all the individual LLM results (Section~\ref{sec:rq1_results}), with improvement ranging from 4.3\% to 15.7\%.  
Following the approach used in the fault localization task, we also experimented with excluding the LLM that had the lowest number of unique correct predictions. 
Excluding LLaMA-3-70B from the combination, still improved the performance over the GPT-3.5 baseline by 9.1\% and all the other individual LLMs. However, the results were lower compared to combining all LLMs. 
This difference in performance can be explained by analyzing the overlap of correct predictions. In LLaMA-3-70B, 58.7\% of the correct predictions were shared among sets of two to five different LLMs, compared to only 37\% for Gemma-7B in fault localization. A set with all six LLMs is excluded as removing one LLM in these cases would likely still yield a correct prediction due to consensus among the remaining LLMs. 
This underscores that in vulnerability detection, LLaMA-3-70B plays a more crucial role in the voting process than Gemma-7B does in fault localization. 
LLaMA-3-70B had more cases where the correct prediction where shared within unique subsets of LLMs.

We found that majority voting has a larger improvement over best-performing LLM in vulnerability detection (3.1\%) compared to fault localization (1.4\%). As shown in Figures~\ref{fig:venn_rq1} and~\ref{fig:venn_rq1_vulnerability}, there are more correctly predicted samples in vulnerability detection that are not shared by all LLMs. Specifically, 71.6\% of correct predictions in vulnerability detection (254/355) come from unique LLM combinations, compared to 59.8\% in fault localization (135/259). The higher percentage of correct predictions shared among unique LLM sets 
in vulnerability detection explains why majority voting is more effective for this task. Additionally, vulnerability detection involves a binary output, which makes majority voting easier. 
These results indicate the need for techniques to combine and refine LLM outputs, especially for tasks with a wider range of possible answers.

\begin{tcolorbox}
{\textbf{RQ2 Findings:} 
We observe that combining results from different LLMs through majority voting can enhance overall performance. In fault localization, the improvement is 13.71\%, while in vulnerability detection, the improvement is 11.2\% compared to the GPT-3.5 baseline.
}
\end{tcolorbox}

\subsection{RQ3: LLMs Cross-Validation Technique}
The results from applying validation prompts in the fault localization are shown in Table~\ref{tab:RQ3_topk}. For the naming convention, \textit{[Left LLM]}$\Leftarrow$\textit{[Right LLM]} means that \textit{[Left LLM]} re-evaluated its results using outputs from \textit{[Right LLM]}. The best performance was achieved by GPT-4o$\Leftarrow$LLaMA-3-70B, which improved the GPT-3.5 baseline by 16.24\%. Following that, the validation prompt result of LLaMA-3-70B$\Leftarrow$GPT-4o also showed a notable improvement, improving the baseline by 13.71\%. These results are higher or at least on par compared to the performance achieved by combining all LLMs through majority voting (Section~\ref{sec:rq2_results}) across all metrics. Furthermore, the cross-validation results of medium-ranked LLMs (3rd and 4th in Top-1 accuracy from RQ1), Mixtral-8x7B and GPT-3.5, also demonstrated improvements over the baseline. However, cross-validation with the low-ranked LLMs (5th and 6th from RQ1), LLaMA-3-8B and Gemma-7B, failed to outperform the GPT-3.5 baseline. This outcome is expected given that the initial performances of LLaMA-3-8B and Gemma-7B were significantly lower compared to GPT-3.5, with a reduction of 12.69\% and 15.22\%, respectively.

\begin{table*}[ht]
\caption{Fault localization results using validation prompt}
\label{tab:RQ3_topk}
\centering
\begin{threeparttable}
\centering
\begin{tabular}{|l|l|r|r|r|r|r|r|}
\hline
\textbf{Grp\hyperref[foot:note4]{*}}                     & \textbf{LLM}                           & \multicolumn{1}{l|}{\textbf{Top-1}} & \multicolumn{1}{l|}{\textbf{Top-2}} & \multicolumn{1}{l|}{\textbf{Top-3}} & \multicolumn{1}{l|}{\textbf{$\Delta$GPT-3.5\hyperref[foot:note3]{\tnote{\S}}}} & \multicolumn{1}{c|}{\textbf{$\Delta$Left\hyperref[foot:note1]{\tnote{\textdagger}}}} & \multicolumn{1}{l|}{\textbf{$\Delta$Right\hyperref[foot:note2]{\tnote{\textdaggerdbl}}}} \\ \hline
\multirow{2}{*}{High} & GPT-4o$\Leftarrow$LLaMA-3-70B                        & \textbf{229}                        & 270                        & 283                        & 16.24\%                     & 3.62\%                          &   4.09\%                      \\ \cline{2-8}  & LLaMA-3-70B$\Leftarrow$GPT-4o                        & 224                        & 265                        & 284                        & 13.71\%                     & 1.82\%                         &     1.36\%                    \\ \hline 
                                 
 \multirow{2}{*}{Med.}                                & GPT-3.5$\Leftarrow$Mixtral-8x7B         & 202                        & 248                        & 271                        & 2.54\%                      & 2.54\%                         & 1.51\%                        \\ \cline{2-8} 
                                 & Mixtral-8x7B$\Leftarrow$GPT-3.5        & 201                        & 248                        & 259                        & 2.03\%                      & 1.01\%                         & 2.03\%                        \\ \hline 
\multirow{2}{*}{Low}                                 & LLaMA-3-8B$\Leftarrow$Gemma-7B  & 160                        & 218                        & 237                        & -18.78\%                    & -4.19\%                        & -6.98\%                       \\ \cline{2-8} 
                                 & Gemma-7B$\Leftarrow$LLaMA-3-8B & 146                        & 173                        & 189                        & -25.89\%                    & -15.12\%                       & -12.57\%                      \\ \hline

\end{tabular}
\begin{tablenotes}[para]
        \footnotesize
        \item[$*$\label{foot:note4}] The ranking of LLMs based on the results from RQ1 grouped into low, medium, and high ranks.
        \item[\textdagger\label{foot:note1}] Relative improvement of the accuracy compared to the LLM that refining the results; \item[\textdaggerdbl\label{foot:note2}] Relative improvement of the accuracy compared to the LLM that is being used as the additional information; 
        \item[\S\label{foot:note3}] Relative improvement of of the accuracy compared to the GPT-3.5 that being used in the previous studies~\cite{zhou2024large, widyasari2020bugsinpy,jiang2024evaluating}.
    \end{tablenotes}
\end{threeparttable}
\end{table*}

\begin{figure}
\centering
    \subfloat[\centering Fault Localization]{{\includegraphics[scale=0.4]{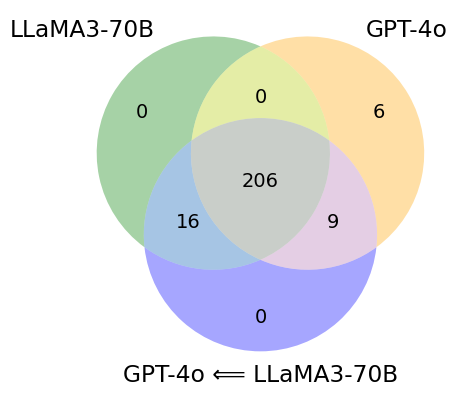} }}%
    \qquad
    \subfloat[\centering Vulnerability Detection]{{\includegraphics[scale=0.4]{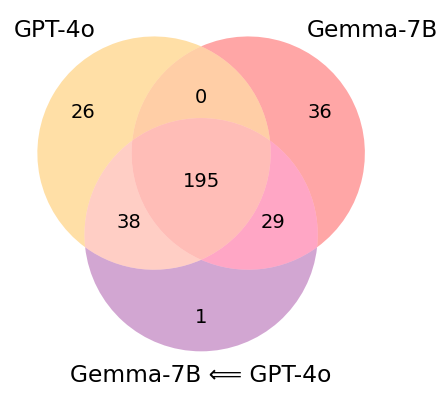}}}%
    \qquad
\caption{Venn diagram of cross-validation prompt result with the default initial prompt.}
\label{fig:venn_rq3_fault_localization}
\end{figure}

From the fault localization results, it appears that Gemma-7B has the least capability to refine its results. It is indicated by the large performance drop of Gemma-7B$\Leftarrow$LLaMA-3-8B  in a 15.12\% and 12.57\% Top-1 performance drop respectively, compared to their initial standalone performances. In contrast,  
the cross-validation results of GPT-4o$\Leftarrow$LLaMA-3-70B showed improvement over running each LLM individually. The Venn diagram in Figure~\ref{fig:venn_rq3_fault_localization}(a) of GPT-4o$\Leftarrow$LLaMA-3-70B, GPT-4o, and LLaMA-3-70B shows that LLaMA-3-70B no longer has uniquely localized fault. There are only 2.7\% of faults (6) were uniquely localized by GPT-4o, with all the other faults also successfully localized by GPT-4o$\Leftarrow$LLaMA-3-70B. In these 6 cases, GPT-4o changes its answer to follow LLaMA-3-70B which makes 5 of them localized at Top-2. 
These findings underscore GPT-4o's effectiveness in enhancing its results through additional insights from another LLM.

\begin{table*}[]
\caption{Vulnerability detection results using validation prompt}
\label{tab:RQ3_vulnerability}
\centering
\addtolength{\tabcolsep}{-0.06em}
\begin{threeparttable}
\begin{tabular}{|l|l|r|r|r|r|r|r|r|r|}
\hline
\textbf{G\hyperref[foot:note4]{*}}                      &\textbf{LLM}                 & \multicolumn{1}{c|}{\textbf{Acc}} & \multicolumn{1}{c|}{\textbf{Prec}} & \multicolumn{1}{c|}{\textbf{Rec}} & \multicolumn{1}{c|}{\textbf{F1}} & \multicolumn{1}{c|}{\textbf{F0.5}}  & \multicolumn{1}{l|}{\textbf{$\Delta$GPT3.5\hyperref[foot:note3]{\tnote{\S}}}} & \multicolumn{1}{c|}{\textbf{$\Delta$Left\hyperref[foot:note1]{\tnote{\textdagger}}}} & \multicolumn{1}{l|}{\textbf{$\Delta$Right\hyperref[foot:note2]{\tnote{\textdaggerdbl}}}} \\ \hline
\multirow{2}{*}{H} & Gemma-7B$\Leftarrow$GPT-4o         & \textbf{67.9}                                   & 70.7                                    & 61.1                                 & 65.6                             & 68.5                               & 8.26\%                        & 0.38\%                & 0.77\%                \\ \cline{2-10}
& GPT-4o$\Leftarrow$Gemma-7B                                        & 67.6                                   & 71.0                                    & 59.6                                 & 64.8                             & 68.4                               & 7.85\%                        & 0.39\%                & 0.00\%                \\ \hline
\multirow{2}{*}{M} & Mixtral-8x7B$\Leftarrow$LLaMA-3-70B & 60.9                                   & 72.8                                    & 34.7                                 & 47.0                             & 59.7                               & -2.90\%                       & -3.69\%               & -4.47\%               \\ \cline{2-10}
& LLaMA-3-70B$\Leftarrow$Mixtral-8x7B            & 63.2                                   & 73.4                                    & 41.5                                 & 53.0                             & 63.6                               & 0.81\%                        & -0.82\%               & 0.00\%                \\ \hline
\multirow{2}{*}{L} & GPT-3.5$\Leftarrow$LLaMA-3-8B       & 60.4                                   & 64.5                                    & 46.1                                 & 53.8                             & 59.7                               & -3.73\%                       & -3.73\%               & -0.85\%               \\ \cline{2-10}
& LLaMA-3-8B$\Leftarrow$GPT-3.5               & 63.0                                   & 74.5                                    & 39.4                                 & 51.5                             & 63.2                               & 0.40\%                        & 3.40\%                & 0.40\%                \\ \hline

\end{tabular}
\end{threeparttable}
\end{table*}
The results of the validation prompts for vulnerability detection are shown in Table~\ref{tab:RQ3_vulnerability}. The highest accuracy was achieved by Gemma-7B$\Leftarrow$GPT-4o, which improved the GPT-3.5 baseline by 8.26\%. Then, it is followed by the GPT-4o$\Leftarrow$Gemma-7B results with a 7.85\% improvement. However, unlike the fault localization task results, these improvements were lower compared to combining all LLMs using majority voting.  Additionally, the validation prompt results for the medium-ranked and low-ranked LLMs, categorized from initial prompt results, only have a minimal improvement of less than 1\% or even a performance decrease of up to 3.73\%. 

Examining the validation prompt results and their individual performances, we found that Gemma-7B$\Leftarrow$GPT-4o showed only a small accuracy improvement (less than 1\%) compared to running Gemma-7B and GPT-4o individually. The Venn diagram for Gemma-7B$\Leftarrow$GPT-4o in vulnerability detection is shown in Figure~\ref{fig:venn_rq3_fault_localization}(b). The results show there are still several codes that were uniquely predicted only by GPT-4o (26) and Gemma-7B (36). This suggests that the refinement process in vulnerability detection is less effective compared to fault localization, where there are generally fewer unique correct predictions by individual LLMs. A notable difference between the prompts in vulnerability detection and fault localization is that the vulnerability detection prompt does not require the LLMs to provide explanations for their results.  
The additional information provided to the corresponding LLM for refinement may only include the binary outcomes (whether it is vulnerable or not) without any reasoning behind it. As a result, the LLM might follow or reject the other LLM's results blindly without understanding the underlying reasoning, potentially leading to less effective refinements.

\begin{tcolorbox}
{\textbf{RQ3 Findings:} 
In fault localization, GPT-4o$\Leftarrow$LLaMA-3-70B improves the performance over the individual model by 4.09\% (LLaMA-3-70B) and 3.62\% (GPT-4o). For vulnerability detection, Gemma-7B$\Leftarrow$GPT-4o results in a smaller improvement (<1\%).
}
\end{tcolorbox}

\subsection{RQ4: Effect of Explanation in LLMs Cross-Validation Technique} 
To address this research question, we reran Gemma-7B and GPT-4o, which showed the best results in previous RQs, using the modified initial prompt by adding a Zero-shot CoT prompt~\cite{kojima2022large} to obtain the reasoning behind the LLM's answers. 
For example, here is a snippet of explanation generated by GPT-4o, for sample 9099: ``\textit{There is a potential SQL Injection vulnerability: the code directly uses the fetched strings in subsequent operations for database login without sanitizing them.}'' This explanation clarifies the nature of the vulnerability and how it is present in the code.
The results of Gemma-7B, GPT-4o, and their cross-validation are highlighted in Table~\ref{tab:RQ4_vulnerability}. We found that the result of GPT-4o$\Leftarrow$Gemma-7B (i.e., GPT-4o refining its answers based on Gemma-7B's information) provided the best outcome, with an improvement of 11.98\% compared to the GPT-3.5 baseline. It also has improvements of 7.13\% and 4.23\% compared to individual runs of GPT-4o and Gemma-7B, respectively. Interestingly, the individual LLMs with explanations showed a slight decrease in performance compared to those without explanations, with decreases of 2.76\% for GPT-4o and 0.35\% for Gemma-7B. However, the cross-validation result of GPT-4o$\Leftarrow$Gemma-7B with explanations was still higher than the validation prompt results using the initial prompt without explanations. This demonstrates better refinement in GPT-4o when explanations are included, compared to without explanations.

Results from running the validation prompt without requiring explanations in RQ3 showed that Gemma-7B$\Leftarrow$GPT-4o achieved the best performance in terms of accuracy compared to other LLM combinations. However, when explanations were required, the results for Gemma-7B$\Leftarrow$GPT-4o indicated that it was more effective to run GPT-4o and Gemma-7B individually. This finding aligns with the fault localization results, where using Gemma-7B to refine results led to a decline in performance compared to running LLMs individually. In 121 cases where GPT-4o and Gemma-7B produced different results, Gemma-7B changed its answer to follow GPT-4o's in 110 cases (90.1\%). In contrast, for cases without explanations, Gemma-7B altered its answers less frequently, at a rate of 42\% (55 out of 129). 
This highlights a limitation of Gemma-7B in refining its results when explanations are included, as it tends to follow the other LLM's answer even if the explanation may not be accurate.   

\begin{table*}[]
\caption{Vulnerability detection results using the Chain-of-Thought prompt and the validation prompt}
\label{tab:RQ4_vulnerability}
\centering
\begin{threeparttable}
\begin{tabular}{|l|r|r|r|r|r|r|r|r|}
\hline
\textbf{LLM}                 & \multicolumn{1}{c|}{\textbf{Acc}} & \multicolumn{1}{c|}{\textbf{Prec}} & \multicolumn{1}{c|}{\textbf{Rec}} & \multicolumn{1}{c|}{\textbf{F1}} & \multicolumn{1}{c|}{\textbf{F0.5}} &  \multicolumn{1}{l|}{\textbf{$\Delta$GPT-3.5\hyperref[foot:note3]{\tnote{\S}}}} & \multicolumn{1}{c|}{\textbf{$\Delta$Left\hyperref[foot:note1]{\tnote{\textdagger}}}} & \multicolumn{1}{l|}{\textbf{$\Delta$Right\hyperref[foot:note2]{\tnote{\textdaggerdbl}}}}\\ \hline
GPT-4o                              & 65.54                                  & 63.76                                   & 72.02                                & 67.64                            & 65.26                              & 4.53\%                        & -                            & -                            \\ \hline
Gemma-7B                              & 67.36                                  & 70.55                                   & 59.59                                & 64.61                            & 68.05                              & 7.43\%                        & -                            & -                            \\ \hline
Gemma-7B$\Leftarrow$GPT-4o & 64.77                                  & 62.67                                   & 73.06                                & 67.46                            & 64.50                              & 3.30\%                        & \multicolumn{1}{r|}{-3.85\%} & \multicolumn{1}{r|}{-1.17\%} \\ \hline
 GPT-4o$\Leftarrow$Gemma-7B  & \textbf{70.21}                                 & 68.93                                   & 73.58                                & 71.18                            & 69.81                              & 11.98\%                       & \multicolumn{1}{r|}{7.13\%}  & \multicolumn{1}{r|}{4.23\%}  \\ \hline
\end{tabular}
\end{threeparttable}
\end{table*}

Compared to the majority voting mechanism using all LLMs in RQ2, the best results from the validation prompt in vulnerability detection show a small improvement of 0.7\% in accuracy. Majority voting requires running six different LLMs, whereas the validation prompt only involves two. This cross-validation approach offers a more efficient solution for enhancing accuracy in vulnerability detection tasks with fewer LLM runs.

\begin{tcolorbox}
{\textbf{RQ4 Findings:} 
Explanations in the initial prompt results impact the LLM refinement process. GPT-4o$\Leftarrow$Gemma-7B with explanations outperforms the version without explanations by 3.8\%, with an improvement of 7.13\% and 4.23\% from the individual runs of respective LLMs.
}
\end{tcolorbox}
\section{Implications}
\label{sec:discussion}
\textbf{Effeciency and Practical Implication.} The validation prompt approach, which involves one LLM refining its results based on another LLM's output, can improve the accuracy of SQA tasks. It offers a cost-effective alternative to the majority voting mechanism that requires running several different LLMs. This is particularly advantageous for tasks where computational resources and costs are a concern. This validation approach can be integrated into existing SQA workflows to enhance the reliability of LLM-based fault localization and vulnerability detection approaches.

\textbf{LLM and Answer Refinement} Our findings demonstrate the capacity of LLMs to refine their outputs. 
If the users aim for better performance compared to a single LLM, we suggest the use of the GPT-4o model to refine the answers. Conversely, the limited refinement capability observed in models like Gemma-7B highlights the need for ongoing monitoring and evaluation of LLMs outputs, especially when these outputs guide critical decisions.

\textbf{Tasks and LLMs Performance.} We observed that LLM can perform differently in different tasks. In the more complex output task of fault localization, Gemma-7B has the worst performance compared to the other LLMs in this study. Meanwhile, in the task with a simpler binary output of vulnerability detection, Gemma-7B excels, achieving the highest accuracy compared to other LLMs. Through this observation, we highlight the importance of analyzing and comparing the results of various LLMs in different tasks.

\section{Threats to Validity}
\label{sec:threats}
An external threat to our study is the generalizability of our findings, as they are based on two specific SQA tasks—fault localization and vulnerability detection—using datasets from previous studies~\cite{widyasari2024demystifying, zhou2024large}. The results for other SQA tasks, such as code review, or when using different datasets may differ from our current findings.
In future work, we plan to evaluate a broader range of tasks and datasets to provide a more comprehensive understanding of LLM performance across various SQA contexts.

ChatGPT and other LLMs used in our study do not provide specific details on the training data, so we lack the means to validate any overlap (i.e., data leakage) with the evaluation dataset. However, it is important to note that the explicit labels for fault location and vulnerability/non-vulnerability methods in the dataset were only introduced in 2024, which is after the latest training data cut-off (Dec. 2023 for LLaMA-3-80B). 
Moreover, the comparison in our study is based on the non-fine-tuned versions of these models, ensuring a fair assessment of their out-of-the-box capabilities without explicitly exposing our specific test cases. 

Given the rapid evolution of LLMs, it is important to note that the output generated by the current prompt may vary across different versions. To mitigate potential threats, we highlight the specific version of the model utilized in this study. Further details regarding the model settings can be found in Section~\ref{sec:llm_setting}. Furthermore, to reduce the effect of the randomness of the experiments, we first run the initial prompts using each LLM 3 times. We found that the result differences between each run are small and the performance ranking of the LLMs does not change across these runs.

Threats to internal validity in our study refer to possible errors in our experiment. To mitigate the risk we have carefully checked the experiments that we run in our study. The code, prompt, and results are also included in our replication package: \url{https://figshare.com/s/5da14b0776750c6fa787}.

\section{Conclusion and Future Work}
\label{sec:conclusion}

In this study, we examined the effectiveness of various LLMs in enhancing SQA tasks. Our research highlighted the distinct capabilities and performance variations among different LLMs, including GPT-3.5, GPT-4o, LLaMA-3-8B, LLaMA-3-70B, Gemma-7B, and Mixtral-8x7B, specifically in the context of automated fault localization and vulnerability detection. We found that LLMs used in this study outperformed all non-LLM-based baseline techniques for both tasks. For fault localization, GPT-4o achieved the best results with a 16.24\% improvement over GPT-3.5. In vulnerability detection, Gemma-7B outperformed the GPT-3.5 baseline by 7.8\%. 
To enhance the accuracy, we employed a voting mechanism among different LLMs, where the results on vulnerability detection and fault localization outperform running the LLMs individually. 

We also proposed a validation prompt method, which facilitates cross-validation of results between LLMs to further refine the outcomes. In fault localization, the result of GPT-4o$\Leftarrow$LLaMA-3-70B (i.e., GPT-4o refining its answers based on LLaMA-3-70B’s output) improves the performance over running GPT-4o (3.6\%) and LLaMA-3-70B (4.1\%) individually. Meanwhile, in vulnerability detection, the result of GPT-4o$\Leftarrow$Gemma-7B improves the performance of running GPT-4o (7.1\%) and LLaMA-3-70B (4.2\%) individually. 
These findings highlighted that while individual LLMs have their own strengths and weaknesses, their integration can further improve their effectiveness in localizing faults and detecting vulnerabilities.

For future work, we plan to investigate a wider range of tasks. We also plan to include the investigation of more LLMs, specifically comparing the general-purpose LLMs with the LLMs that are designed for the specific task in SQA (e.g., LLMAO~\cite{yang2024large} for fault localization).
\bibliographystyle{ACM-Reference-Format}
\bibliography{custom}


\end{document}